\documentclass[twoside, 11pt]{article}

\usepackage{blindtext}
\usepackage{multirow}
\usepackage{tabularx}
\usepackage{array}
\usepackage[table]{xcolor}
\usepackage{listings}
\usepackage{adjustbox}
\usepackage{arydshln}
\usepackage{caption}
\usepackage{float}
\usepackage{enumitem}
\usepackage{comment}
\usepackage{floatflt}
\usepackage{wrapfig}
\usepackage{booktabs}
\usepackage{tcolorbox}
\usepackage{listings}
\usepackage{colortbl}
\usepackage{fontawesome}
\usepackage{url}
\usepackage{footnote}  
\usepackage{titletoc}
\usepackage[toc,page,header]{appendix}
\usepackage{minitoc}

\usepackage{threeparttable}

\usepackage{makecell}

\definecolor{codegreen}{rgb}{0,0.6,0}
\definecolor{backcolour}{rgb}{0.97,0.97,0.97}

\lstdefinestyle{bashstyle}{ backgroundcolor=
\color{backcolour}
, basicstyle=\ttfamily\footnotesize, breaklines=true, showstringspaces=false,
keywordstyle=
\color{blue}
, commentstyle=
\color{codegreen}
, frame=none, numbers=none, }

\usepackage{jmlr2e}

\usepackage{lastpage}
\jmlrheading{26}{2025}{1-\pageref{LastPage}}{7/31}{7/25}{21-0000}{Leyi Pan, Sheng Guan, Zheyu Fu, Luyang Si, Zian Wang, Xuming Hu, Irwin King, Philip S.Yu, Aiwei Liu, Lijie Wen}

\ShortHeadings{\textsc{MarkDiffusion}: An Open-Source Toolkit for Generative Watermarking of Latent Diffusion Models}{Pan, Guan, Fu, Si, Wang, Hu, King, Yu, Liu, Wen}
\firstpageno{1}

\begin{document}
  \title{\textsc{MarkDiffusion}: An Open-Source Toolkit for Generative
  Watermarking of Latent Diffusion Models}

  \author{\name Leyi Pan$^{1*}$
  \email{panly24@mails.tsinghua.edu.cn}
  \\
  \name Sheng Guan$^{1,2*\ddagger}$
  \email{guansheng2022@bupt.edu.cn}
  \\
  \name Zheyu Fu$^{1*}$
  \email{fuzy23@mails.tsinghua.edu.cn}
  \\
  \name Luyang Si$^{1*}$
  \email{sily23@mails.tsinghua.edu.cn}
  \\
  \name Huan Wang$^{1}$
  \email{huan-wan23@mails.tsinghua.edu.cn}
  \\
  \name Zian Wang$^{1}$
  \email{arthurwzaa@gmail.com}
  \\
  \name Hanqian Li$^{5}$
  \email{hli994@connect.hkust-gz.edu.cn}
  \\
  \name Xuming Hu$^{5}$
  \email{xuminghu97@gmail.com}
  \\
  \name Irwin King$^{3}$
  \email{king@cuhk.edu.hk}
  \\
  \name Philip S. Yu$^{4}$
  \email{psyu@uic.edu}
  \\
  \name Aiwei Liu$^{1\dagger}$
  \email{liuaiwei20@gmail.com}
  \\
  \name Lijie Wen$^{1\dagger}$
  \email{wenlj@tsinghua.edu.cn}
  \\
  $^{1}$\textsc{Tsinghua University} \quad $^{2}$\textsc{Beijing University of
  Posts and Telecommunications} \\
  $^{3}$\textsc{The Chinese University of Hong Kong} \quad $^{4}$\textsc{University of Illinois at Chicago} \\
  $^{5}$\textsc{The Hong Kong University of Science and Technology (Guangzhou)}\\
  {\faGithub\ \textcolor{magenta}{[Official]: \url{https://github.com/THU-BPM/MarkDiffusion}}}
  }
  
  \editor{Editor}

  \maketitle

  \makeatletter \def\@fnsymbol#1{\ensuremath{\ifcase#1\or \ddagger\or *\or \dagger\else\@ctrerr\fi}}
  \makeatother

  \renewcommand{\thefootnote}{\fnsymbol{footnote}}
  \setcounter{footnote}{0}

  \stepcounter{footnote}
  \footnotetext{Work done during internship at Tsinghua University.}
  \stepcounter{footnote}
  \footnotetext{Equal contribution.}
  \stepcounter{footnote}
  \footnotetext{Corresponding authors.}

  \begin{abstract}
    We introduce \textsc{MarkDiffusion}, an open-source Python toolkit for generative watermarking of latent diffusion models. It comprises three key components: a \texttt{unified implementation framework} for streamlined watermarking algorithm integrations and user-friendly interfaces; a \texttt{mechanism visualization suite} that intuitively showcases added and extracted watermark patterns to aid public understanding; and a \texttt{comprehensive evaluation module} offering standard implementations of 24 tools across three essential aspects—detectability, robustness, and output quality—plus 8 automated evaluation pipelines. Through \textsc{MarkDiffusion}, we seek to assist researchers, enhance public awareness and engagement in generative watermarking, and promote consensus while advancing research and applications.
  \end{abstract}

  \begin{keywords}
    watermark, latent diffusion models, toolkits, evaluation, visualization
  \end{keywords}

\section{Introduction}
With the advancement of diffusion models~\citep{ho2020denoising, song2020denoising}, Latent Diffusion Models (LDMs) such as Stable Diffusion~\citep{rombach2021highresolution} have excelled in generating high-quality images and videos, with broad applications in digital art and film production. However, these capabilities also raise risks of misuse, such as misleading public opinion or infringing on intellectual property~\citep{bird2023typology, jaidka2025misinformation}. 

This has driven the development of watermarking techniques, which embed imperceptible signals detectable by algorithms to identify machine-generated content~\citep{wenTreeRingWatermarksFingerprints2023,fernandez2023stable}. Generative watermarking, specifically, injects signals into the latent space during LDM inference, requiring no model retraining or post-processing, making it the dominant approach. As an emerging technology, generative watermarking for LDMs faces key challenges: (i) incompatible and hard-to-reuse algorithm implementations; (ii) high technical complexity impeding understanding and experimentation; and (iii) insufficient comprehensive evaluation frameworks for performance assessment. To tackle these, we present \textsc{MarkDiffusion}, an open-source Python toolkit comprising:
\begin{itemize}[itemsep=0pt, parsep=0pt, topsep=0pt, partopsep=0pt]
  \item A unified implementation framework that seamlessly integrates eight state-of-the-art LDM watermarking algorithms with modular design and easy-to-use APIs.
  \item A comprehensive evaluation module that integrates 24 tools and 8 automated pipelines, covering detectability, robustness, and output quality.
  \item An interactive visualization suite that intuitively demonstrates watermark embedding and detection mechanisms, enhancing accessibility for a wider audience.
\end{itemize}

\begin{figure}[t]
  \centering
  \vspace{-1cm}
  \includegraphics[width=0.99\textwidth]{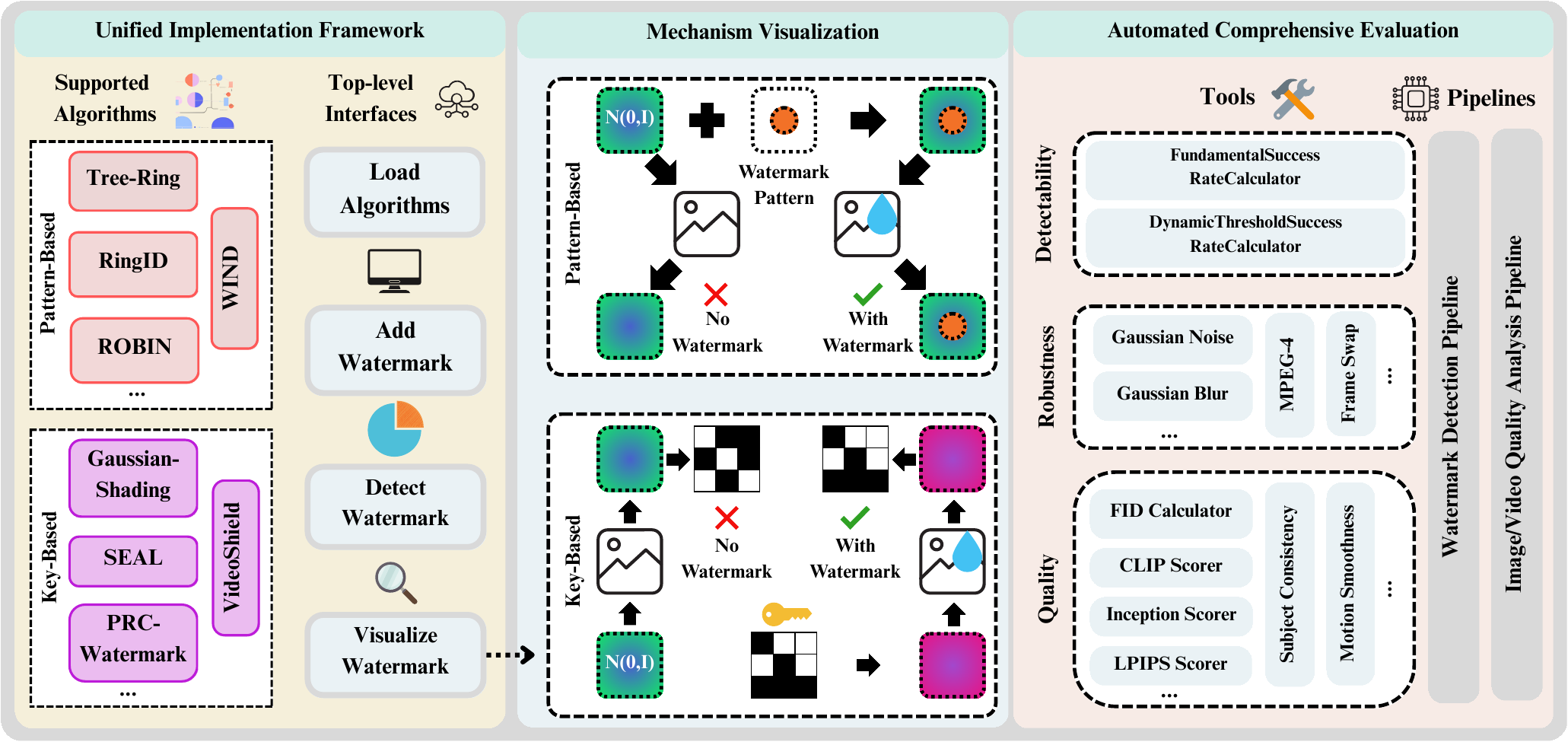}
  \vspace{-0.3cm}
  \caption{Architecture overview of \textsc{MarkDiffusion}.}
  \label{fig:ring_id_visualization}
  \vspace{-0.5cm}
\end{figure}
  \section{Toolkit Design}

\subsection{Unified Implementation Framework}

The design of implementation framework offers the following three key advantages. 

        

\begin{wraptable}{t}{0.5\textwidth}
    \vspace{-22pt}
    \centering
    \captionof{table}{\small Supported watermarking algorithms.}
    \label{tab:watermarking_algorithms}
    \scriptsize
    \begin{tabularx}{0.5\textwidth}{l c c}
        \toprule
        \textbf{Algorithms} & \textbf{Cat.} & \textbf{Target} \\
        \midrule
        Tree-Ring~\citep{wenTreeRingWatermarksFingerprints2023} & Pat. & Image \\
        Ring-ID~\citep{ciRingIDRethinkingTreeRing2024} & Pat. & Image \\
        ROBIN~\citep{huangROBINRobustInvisible2025} & Pat. & Image \\
        WIND~\citep{arabiHiddenNoiseTwoStage2025} & Pat. & Image \\
        Gaussian-Shading~\citep{yangGaussianShadingProvable2024} & Key & Image \\
        PRC~\citep{gunnUndetectableWatermarkGenerative2024} & Key & Image \\
        SEAL~\citep{arabiSEALSemanticAware2025} & Key & Image \\
        VideoShield~\citep{huVideoShieldRegulatingDiffusionbased2025} & Key & Video \\
        \bottomrule
    \end{tabularx}
    \label{tab:algorithms}
    \vspace{-2pt}
\end{wraptable}

\noindent\textbf{Comprehensive Coverage.} As shown in Table~\ref{tab:algorithms}, our toolkit implements eight state-of-the-art watermarking algorithms spanning two mainstream categories: (i) \emph{Pattern-Based methods}, which embed predefined patterns (e.g., rings in Fourier space) into the initial noise of LDM inference and detect them by measuring pattern similarity post-inversion; and (ii) \emph{Key-Based methods}, which use watermark keys to generate initial noise and detect via comparison with recovered noise. Both image and video watermarking algorithms are supported. Details of module and class design are in Appendix~\ref{appendix:module}.

\begin{figure}[t]
    \centering
    \includegraphics[width=0.95\linewidth]{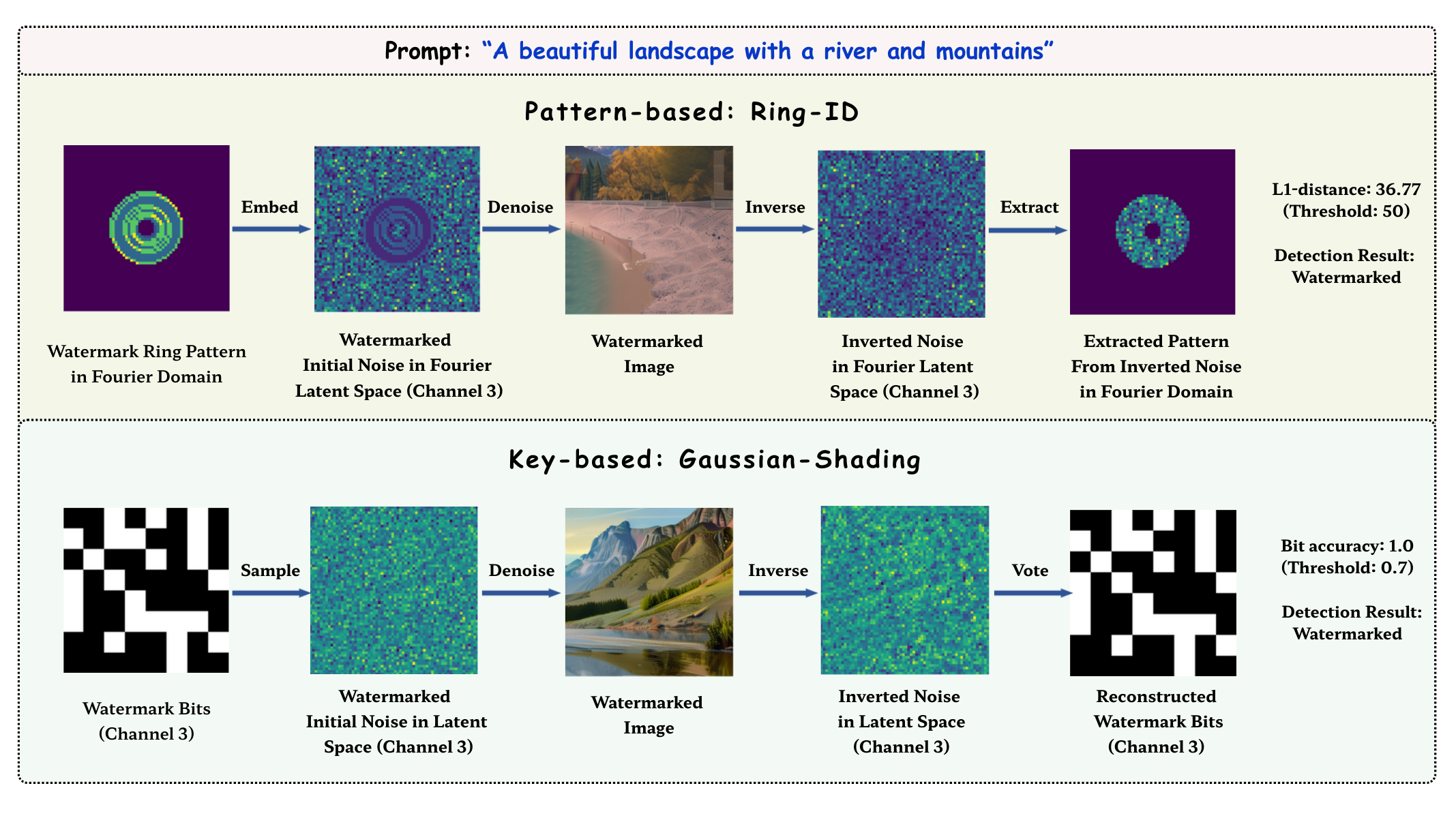}
    \vspace{-0.4cm}
    \caption{Visualization examples using \textsc{MarkDiffusion}.}
    \label{fig:watermarking_vis_mechanisms}
    \vspace{-0.5cm}
\end{figure}

\begin{wraptable}{t}{0.54\textwidth}
  \vspace{-0.4cm}
  \centering
  \scriptsize
  \caption{\small Taxonomy of the evaluation module in \textsc{MarkDiffusion}.}
  \vspace{-0.2cm}
  \label{tab:evaluation_tools}
  
  \scriptsize
  \begin{tabular}{p{0.10\textwidth} p{0.25\textwidth} p{0.12\textwidth}}
  \toprule
  \textbf{Perspective} & \textbf{Tools} & \textbf{Pipeline} \\
  \midrule
 \multirow{2}{*}{Detectability} 
    & Fundamental SR. Calculator, & UWMDetect, \\ 
    & Dynamic SR. Calculator & WMDetect \\
\midrule
  \multirow{4}{*}{\makecell{Robustness \\ (Image)}} 
      & Rotation, Crop \& Scale, & \multirow{4}{*}{\makecell{UWMDetect, \\ WMDetect}} \\
      & Gaussian Noise, & \\ 
      & Gaussian Blur, & \\
      & JPEG, Color Jitter& \\  
  \midrule
  \multirow{3}{*}{\makecell{Robustness \\ (Video)}} 
      & MPEG-4, & \multirow{3}{*}{\makecell{UWMDetect, \\ WMDetect}} \\
      & Frame Average, & \\
      & Frame Swap & \\  
  \midrule
  \multirow{5}{*}{\makecell{Quality \\ (Image)}} 
      & Inception Score, FID, & GroupQual, \\
      & LPIPS, & RepeatQual, \\
      & CLIP-I, CLIP-T, & RefQual, \\
      & PSNR, SSIM, & CompQual, \\
      & NIQE & DirectQual \\
  \midrule
  \multirow{5}{*}{\makecell{Quality \\ (Video)}} 
      & Subject Consistency, & \multirow{5}{*}{VideoQual} \\
      & Background Consistency, & \\
      & Motion Smoothness, & \\
      & Dynamic Degree, & \\  
      & Imaging Quality & \\  
  \bottomrule
  \end{tabular}
  \vspace{-0.3cm}
\end{wraptable}


\noindent\textbf{Modular Extensibility.} \textsc{MarkDiffusion} decouples the
watermark pipeline into three independent modules: watermark embedding, watermark detection, and latent inversion. This modular design allows developers to easily extend the toolkit with new algorithms or customize existing components. For instance, users can plug in alternative detection metrics (e.g., p-value) or configure different inversion algorithms (e.g., DDIM Inversion~\citep{song2020denoising} or Exact Inversion~\citep{hongExactInversionDPMSolvers2023}) without modifying the core framework.

\vspace{3pt}

\noindent
\textbf{User-Friendly Interface.} All algorithms can achieve watermark embedding and detection through direct calls to \texttt{generate\_watermarked\_media()}
and \texttt{detect\_watermark\_in\_media()}, respectively. Moreover, all algorithms are fully compatible with the latest \texttt{Diffusers} library. Comprehensive user cases are provided in Appendix~\ref{appendix:user-cases}.

\subsection{Visualization Suite}
\textsc{MarkDiffusion} provides tailored visualization modules for each algorithm, intuitively demonstrating how watermarks are embedded, extracted, and verified. As illustrated in Figure~\ref{fig:watermarking_vis_mechanisms}, (i) for pattern-based methods like Ring-ID, the visualization clearly shows that both the embedded and extracted patterns exhibit a ring shape, with their L1-distance falling below the preset threshold; (ii) for key-based methods such as Gaussian-Shading, it displays the initial watermark key, the sampled initial noise derived from it, and the extracted watermark bits, intuitively highlighting their similarity. Appendix~\ref{appendix:visualize} provides additional visualization examples.

\vspace{-0.3cm}

\subsection{Evaluation Module}
Evaluating watermarking algorithms for LDMs requires assessment across three essential dimensions: (1) \textbf{Watermark Detectability}, which measures the ability to reliably distinguish watermarked from non-watermarked content; (2) \textbf{Robustness Against Attacks}, which tests detectability after common manipulations such as compression, cropping, or noise addition; and (3) \textbf{Impact on Media Quality}, which quantifies how watermarking affects the quality of generated images or videos. To address these, \textsc{MarkDiffusion} provides standard implementations of up to 24 evaluation tools, along with 8 automated pipelines that integrate steps such as watermarked content generation, manipulations, watermark extraction, and metric computation into seamless workflows, as detailed in Table~\ref{tab:evaluation_tools}.
  \section{Experiments}
We validate \textsc{MarkDiffusion}'s ability to reproduce experimental results through comprehensive evaluations. All experiments use Stable Diffusion v2.1~\citep{rombach2021highresolution} for image generation and ModelScope~\citep{wang2023modelscope} for video generation. For image content, we utilize 200 prompts from the Stable Diffusion Prompt dataset~\citep{sdp_dataset}, while video content prompts are drawn from VBench~\citep{huangVBenchComprehensiveBenchmark2023}. Table~\ref{tab:comprehensive_evaluation} compares the test results between implementations using our toolkit (x/) and the official implementation (/x). Our toolkit effectively reproduces the outcomes of all experiments.

\vspace{-0.5cm}

\begin{table}[h]
    \centering
    \caption{Evaluation results of watermarking algorithms across key metrics.}
    \label{tab:comprehensive_evaluation}
    \small
    \setlength{\tabcolsep}{4pt}
    \resizebox{\textwidth}{!}{
    \begin{tabular}{ccccccc}
        \toprule
        \multirow{2}{*}{\textbf{Algorithms}} & 
        \multirow{2}{*}{\makecell[cc]{\textbf{Detectability($\uparrow$)} \\ \textbf{(TPR@FPR=1\%)}}}  & 
        \multicolumn{3}{c}{\textbf{Robustness (F1-score$\uparrow$)}} & 
        \multicolumn{2}{c}{\textbf{Quality}} \\
        \cmidrule(l){3-5} \cmidrule(l){6-7}
         &  & \textbf{JPEG} & \textbf{Gaussian Blur} & \textbf{Gaussian Noise} & \textbf{FID($\downarrow$)} & \textbf{CLIP-T($\uparrow$)} \\
        \midrule
        Tree-Ring & 1.00 / 1.00  & 0.85 / 0.88 & 0.66 / 0.68 & 0.95 / 0.93 & 32.30 / 33.15 & 0.658 / 0.665 \\ 
        Ring-ID   & 1.00 / 1.00 & 0.98 / 1.00 & 0.66 / 0.68 & 1.00 / 1.00  & 37.24 / 38.17 & 0.658 / 0.663 \\
        ROBIN     & 1.00 / 1.00 & 0.99 / 1.00 & 1.00 / 1.00  & 0.98 / 1.00 & 23.15 / 25.23 & 0.655 / 0.660 \\
        G-Shading & 1.00 / 1.00 & 0.98 / 0.98  & 0.76 / 0.82 & 0.93 / 0.95 & 35.97 / 36.23 & 0.677 / 0.682 \\
        PRC       & 1.00 / 1.00 & 0.97 / 0.98  & 0.02 / 0.00  & 1.00 / 1.00  & 35.94 / 36.43 & 0.660 / 0.665 \\
        WIND      & 1.00 / 1.00 & 1.00 / 0.99  & 1.00 / 1.00  & 1.00 / 0.98  & 41.50 / 37.91 & 0.650 / 0.659 \\
        SEAL      & 0.95 / 0.98 & 0.79 / 0.81 & 0.67 / 0.67 & 0.68 / 0.67 & 37.35 / 36.97 & 0.659 / 0.664 \\
        \midrule

        \textbf{Algorithms} & 
        \textbf{Detectability($\uparrow$)} & 
        \textbf{MPEG-4} & 
        \textbf{Frame Average} &
        \textbf{Frame Swap} &
        \multicolumn{2}{c}{\textbf{Avg. of 5 Quality Metrics($\uparrow$)}} \\

        \midrule

        VideoShield & 1.00 / 1.00 & 0.98 / 0.98 & 0.99 / 1.00 & 1.00 / 1.00 & \multicolumn{2}{c}{0.803 / 0.804 } \\
        \bottomrule
    \end{tabular}
    }
\end{table}

\vspace{-0.5cm}

  \section{Conclusion}
This paper introduces \textsc{MarkDiffusion}, an open-source toolkit for generative watermarking of Latent Diffusion Models. \textsc{MarkDiffusion} offers a comprehensive suite of eight state-of-the-art watermarking algorithms with a unified, modular, and extensible interface. It also includes a visualization suite and a comprehensive evaluation module featuring up to 24 tools and 8 automated pipelines, enabling researchers to easily compare algorithms and perform systematic experiments.









  
  \bibliography{sample}

  \newpage

  \appendix

  \begin{center}
    {\huge \bfseries Appendices}
    \end{center}
    
    \vspace{3em}
    
    \noindent{\Large \bfseries Table of Contents}
    
    \vspace{0.5em}
    
    \hrule height 0.5pt  
    
    \startcontents[appendix]
    \printcontents[appendix]{}{1}{}
    
    \vspace{0.7em}
    
    \hrule height 0.5pt  
    
    \newpage

  \section{Module and Class Design}
\label{appendix:module}
This section details the system architecture, core modules and class design principles underlying the toolkit.

\subsection{Unified Implementation Framework}
\begin{figure}[t]
    \centering
    \includegraphics[width=\textwidth]{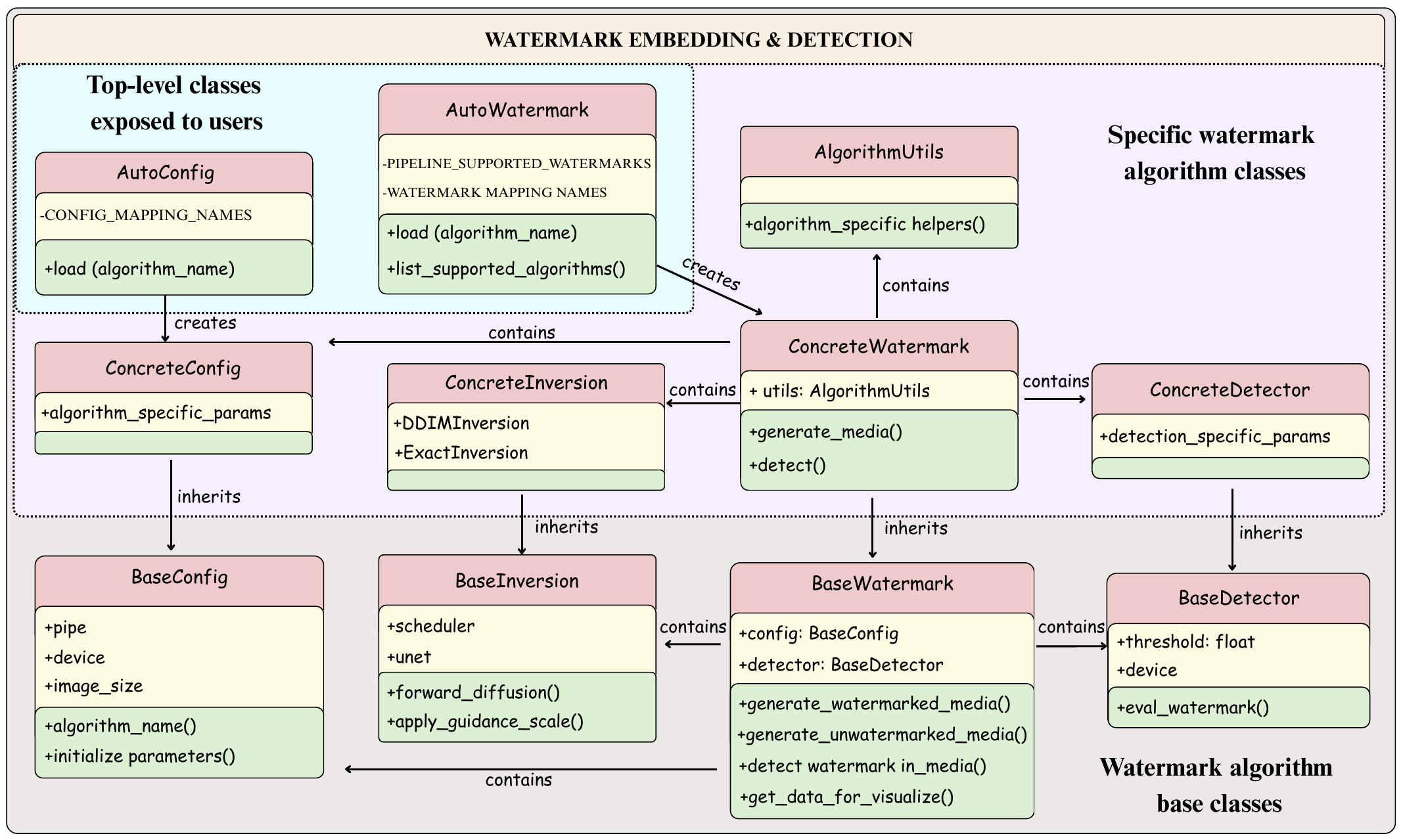}
    \caption{Detailed architecture of the unified implementation framework.}
    \label{fig:structure}
\end{figure}
Figure \ref{fig:structure} illustrates the design architecture of the unified implementation framework module in \textsc{MarkDiffusion}, which is responsible for watermark embedding and detection for each algorithm. The framework adopts a modular design, divided into three layers:
\begin{itemize}
\item \textbf{Top-level classes exposed to users:} To facilitate user interaction, we provide two top-level classes, \texttt{AutoConfig} and \texttt{AutoWatermark}. By using the \texttt{load()} function, users can specify the algorithm name to automatically load the corresponding specific algorithm class (referred to as \texttt{ConcreteWatermark}) and specific configuration class (referred to as \texttt{ConcreteConfig}), as illustrated in the figure.
\item \textbf{Specific watermark algorithm classes:} Each watermark algorithm has a core Watermark class (referred to as \texttt{ConcreteWatermark}). The data members of this class include a \texttt{Config} object, a \texttt{Utils} object, an \texttt{Inversion} object, and a \texttt{Detector} object, each responsible for their respective tasks. The top-level interface of the class includes \texttt{generate\_watermarked\_media()} and \texttt{detect\_watermark\_in\_media()}, which enable watermark embedding and detection with a single line of code.
\item \textbf{Watermark algorithm base classes:} These are the fundamental base classes at the lowest level, from which all specific watermark algorithm classes and their component classes inherit.
\end{itemize}

\subsection{Mechanism Visualization}
\begin{figure}[h!]
    \centering
    \includegraphics[width=1\textwidth]{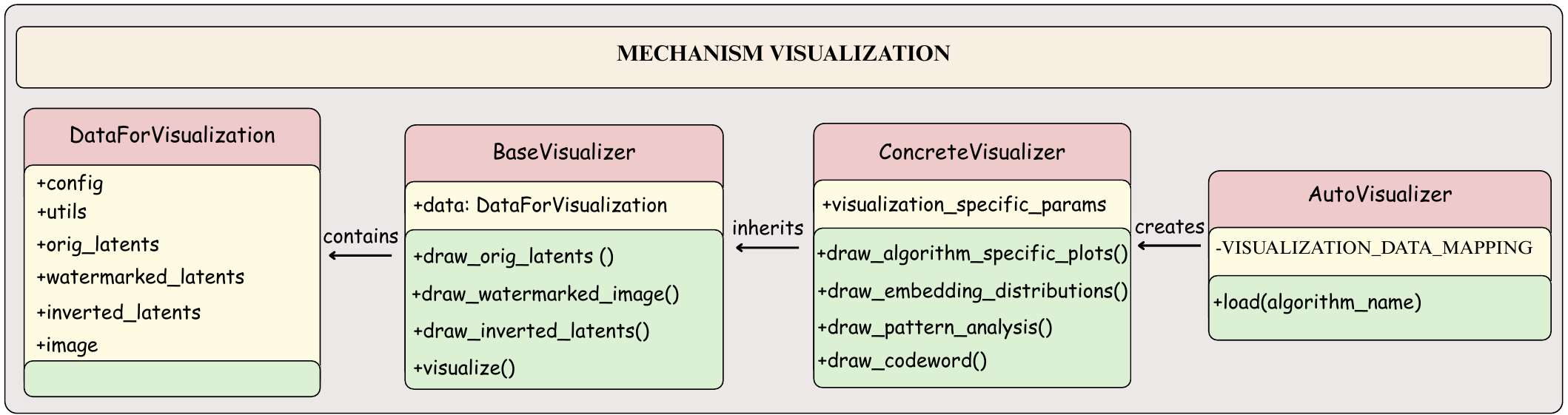}
    \caption{Detailed architecture of the mechanism visualization module.}
    \label{fig:visualization_structure}
\end{figure}

The visualization module enables easy understanding of watermarking mechanisms through algorithm-specific visualizers, as shown in Figure \ref{fig:visualization_structure}. The visualization classes obtain the \texttt{DataForVisualization} data from the algorithm class via the \texttt{get\_data\_for\_visualization()} method. Each specific visualizer (created automatically from the \texttt{load()} method in \texttt{AutoVisualizer}) inherits from \texttt{BaseVisualizer} and implements methods for displaying latent representations, watermark patterns, and frequency-domain characteristics.

\subsection{Evaluation Module}
\begin{figure}[h!]
    \centering
    \includegraphics[width=1\textwidth]{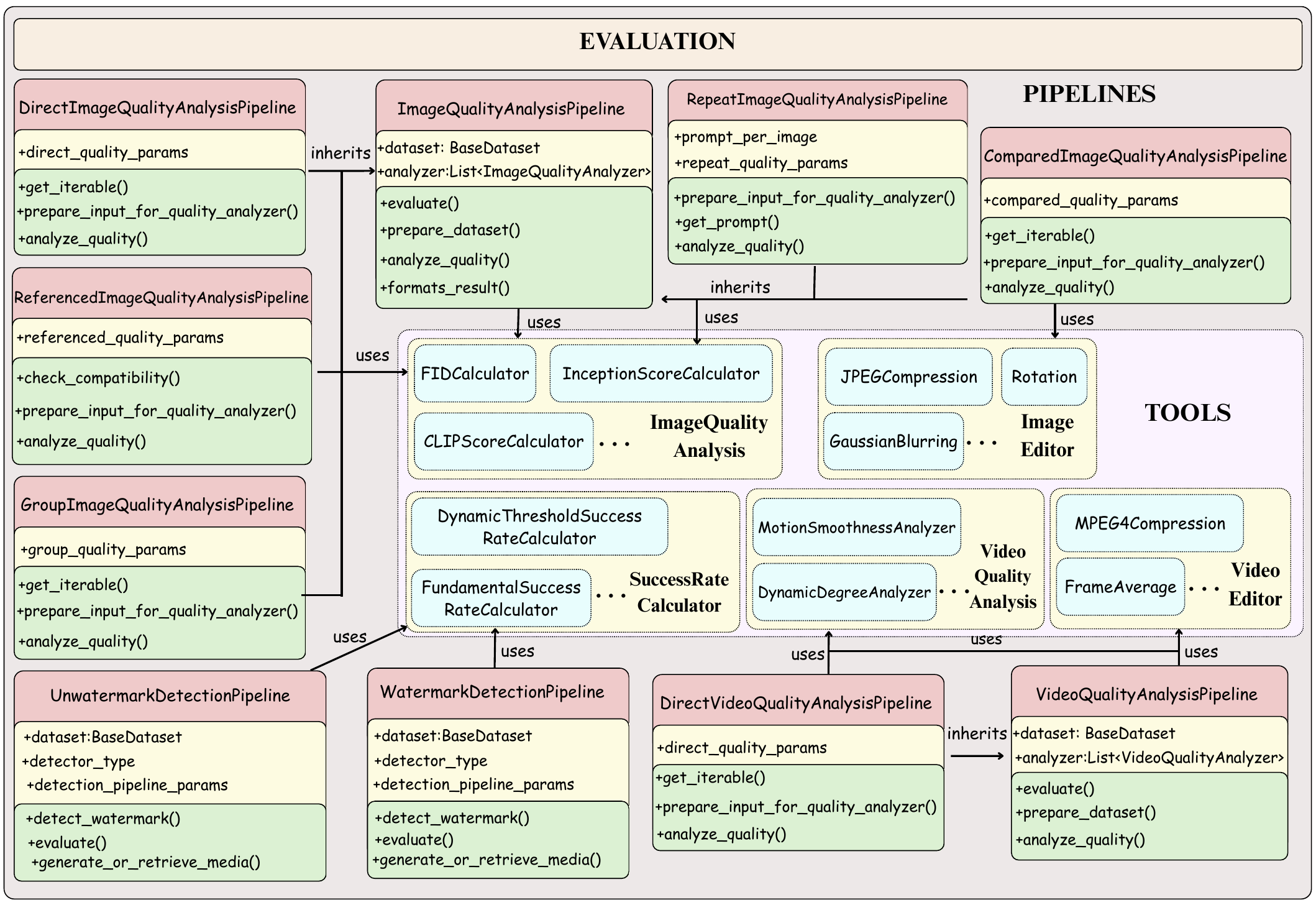}
    \caption{Detailed Structure of the evaluation module.}
    \label{fig:evaluation_structure}
\end{figure}

Figure \ref{fig:evaluation_structure} details the structure of the evaluation modules. The evaluation modules in \textsc{MarkDiffusion} consist of 8 pipelines, covering three evaluation perspectives: detectability, robustness, and impact on media quality. Detectability and robustness are evaluated using the \texttt{DetectionPipeline}, while the impact on quality is assessed through multiple \texttt{QualityPipeline} instances. Each evaluation pipeline can integrate various toolsets.

\subsection{Class Design Principles}
The toolkit architecture is built upon well-established object-oriented design principles, incorporating key design patterns to ensure flexibility, consistency, and maintainability:
\begin{itemize}[itemsep=0.5pt]
    \item \textbf{Template Method Pattern}: The \texttt{BaseWatermark} class defines the algorithmic framework for watermark operations. Subclasses implement specific steps, ensuring a consistent overall structure while allowing customization of individual components.
    \item \textbf{Strategy Pattern}: Detection algorithms and inversion methods are implemented as interchangeable strategies. This design enables dynamic selection and comparison of different approaches during runtime, fostering modularity and extensibility.
    \item \textbf{Factory Pattern}: Both \texttt{AutoWatermark} and \texttt{AutoVisualizer} implement factory mechanisms to automatically instantiate algorithm-specific classes based on string identifiers. This simplifies object creation and reduces code coupling.
    \item \textbf{Composition Over Inheritance}: Complex functionalities are achieved by composing simpler, reusable components rather than relying on deep inheritance hierarchies. This approach enhances code maintainability, improves testing efficiency, and avoids the limitations of rigid inheritance structures.
\end{itemize}
  \section{Background: Generative Watermarking and Post-hoc Watermarking}
\label{appendix:background}
Existing watermarking techniques for diffusion models can broadly be divided into two categories: \textit{generative watermarking}, which integrates watermarking signals directly into the generation process, and \textit{post-hoc watermarking}, which modifies an image or video after it has been generated.

Post-hoc watermarking methods typically utilize traditional digital watermarking techniques, with frequency-domain approaches being particularly prominent. These methods operate by altering representations in the frequency domain using transformations such as the Discrete Wavelet Transform (DWT)~\citep{xia1998wavelet} 
or Discrete Cosine Transform (DCT)~\citep{cox2008digital}. 
A practical example of this is the DwtDct Watermarking method~\citep{navas2008dwt}, which has been applied in the open-source Stable Diffusion model. Frequency-domain techniques are known for their robustness against operations such as cropping, resizing, and compression. 

In addition to these traditional methods, deep encoder-decoder frameworks, such as HiDDeN~\citep{zhu2018hidden}, adopt an end-to-end approach for embedding and extracting watermarks. RivaGAN~\citep{zhang2019robust} extends this framework by incorporating adversarial training, which introduces perturbations and image processing into the model training process to enhance robustness. However, recent studies indicate that this watermarking paradigm lacks resilience against regeneration attacks using advanced generative models~\citep{zhao2024invisible}. Furthermore, post-hoc methods introduce pixel-level perturbations that inevitably degrade the quality of generated images or videos, making it challenging to maintain undetectable watermarks.

Conversely, generative watermarking methods embed watermarking signals directly within the image or video generation process. By avoiding direct modifications to the generated content, these approaches minimize perceptual impacts on the output~\citep{arabiSEALSemanticAware2025} while improving robustness against regeneration attacks. Early generative watermarking techniques incorporate watermarks by embedding them into the training data and fine-tuning specific model components. A representative example of this is Stable Signature~\citep{fernandez2023stable} is a representative. However, these methods treat watermarking as independent from the generation task and require significant model retraining to incorporate the watermark~\citep{huangROBINRobustInvisible2025}, 
which limits their flexibility. 

An alternative approach involves modifying latent representations during the inference stage of latent diffusion models. This allows watermarks to be embedded into generated images or videos without requiring additional training or direct manipulation of the generated content. To detect such watermarks, the latent representations are reconstructed using diffusion inversion algorithms, such as DDIM Inversion~\citep{song2020denoising}, to verify whether the watermarking key remains detectable. An example of this approach is Tree-Ring~\citep{wenTreeRingWatermarksFingerprints2023}, which embeds a Fourier-domain ring pattern into the initial latent space for watermarked image generation. The watermark is then detected by calculating the L1 distance between the reconstructed and original patterns in Fourier space.

\textsc{MarkDiffusion} builds upon this emerging class of generative watermarking techniques by introducing a comprehensive toolkit for the generation, detection, and evaluation of such watermarks. It offers a flexible, modular platform that supports various generative watermarking algorithms, enabling research and experimentation with this technology within Latent Diffusion Models.
  \section{Comparison with Existing Toolkits}
\label{appendix:comparison}
\setcounter{footnote}{0}  
\renewcommand{\thefootnote}{\arabic{footnote}}  

\begin{table}[t]
\centering
\footnotesize
\caption{Comparison of \textsc{MarkDiffusion} with existing watermarking toolkits.}
\label{tab:toolkit_comparison_transposed}
\resizebox{\textwidth}{!}{
\begin{tabular}{l p{0.12\textwidth} p{0.22\textwidth} p{0.36\textwidth}}
\toprule
& \textbf{Watermark Paradigm} & \textbf{Application Domain} & \textbf{Core Focus} \\
\midrule

invisible-watermark & 
Post-hoc & 
General Image Files & 
A lightweight tool for applying watermarks to existing images. \\
\addlinespace

MarkLLM & 
Generative & 
Large Language Models (Text) & 
A comprehensive toolkit for generative text watermarking. \\
\addlinespace 

\rowcolor{gray!20}
\textsc{MarkDiffusion} & 
Generative & 
Diffusion Models (Images \& Video) & 
A comprehensive toolkit for generative watermarking in visual media. \\
\bottomrule
\end{tabular}
}
\end{table}

To situate \textsc{MarkDiffusion} within the broader landscape of watermarking technologies, this section provides a comparative analysis with two relevant toolkits: \textbf{invisible-watermark}\footnote{\url{https://github.com/ShieldMnt/invisible-watermark}} \citep{inv_wm} and \textbf{MarkLLM}\footnote{\url{https://github.com/THU-BPM/MarkLLM}}~\citep{pan-etal-2024-markllm}. This comparison clarifies the distinct technical paradigm of \textsc{MarkDiffusion} and highlights its specific contributions to the field of generative media. The core distinctions are summarized in Table \ref{tab:toolkit_comparison_transposed}.

The primary distinction between \textsc{MarkDiffusion} and invisible-watermark lies in their fundamental watermarking paradigm. invisible-watermark operates as a post-hoc tool, designed to apply watermarks to images after the generation process is complete. In contrast, \textsc{MarkDiffusion} implements a generative or in-process approach, integrating the watermarking procedure directly into the synthesis pipeline of diffusion models for both images and videos. This methodological divergence means \textsc{MarkDiffusion} embeds provenance information at the point of creation, rather than as a subsequent modification.

While \textsc{MarkDiffusion} shares a generative paradigm with MarkLLM, their application domains are entirely different. MarkLLM is a pioneering framework developed for watermarking content generated by Large Language Models (LLMs), focusing exclusively on the text modality~\citep{DBLP:conf/icml/KirchenbauerGWK23,christ2023undetectable,liu2024a,liu2024an,liu2025can,pan2025can,pan2025waterseeker,textwatermarksurvey}. Nevertheless, MarkLLM provided significant conceptual inspiration for \textsc{MarkDiffusion}. Specifically, its architectural philosophy, including the Python package structure and the design of its evaluation pipeline, served as a foundational reference. \textsc{MarkDiffusion} adapts and extends these principles to meet the unique challenges and requirements of the visual domain, thereby bridging the gap between generative watermarking concepts in text and their practical implementation in image and video synthesis.

In summary, \textsc{MarkDiffusion} is not positioned as a replacement for these tools but rather as a solution that addresses a distinct and critical need. It diverges from the post-hoc methodology of invisible-watermark by championing an in-process approach. Furthermore, it effectively translates and adapts the generative watermarking philosophy established by MarkLLM in the text domain to the visual media landscape, offering a specialized and comprehensive toolkit for diffusion models.
  \section{User Cases}
\label{appendix:user-cases}
 The following code snippets demonstrate examples
 of how to use MarkDiffusion in one’s project. 
\subsection{Cases for Generating and detecting Watermarked Media}
\begin{lstlisting}[language=Python]
# Load watermarking algorithm
mywatermark = AutoWatermark.load(
    'GS', 
    algorithm_config=f'config/GS.json',
    diffusion_config=diffusion_config
)
# Generate watermarked image
watermarked_image = mywatermark.generate_watermarked_media(
    input_data="A beautiful landscape with a river and mountains"
)
# Visualize the watermarked image
watermarked_image.show()
# Detect watermark
detection_result = mywatermark.detect_watermark_in_media(watermarked_image)
print(detection_result)
\end{lstlisting}
\subsection{Cases for Visualizing Watermarking Mechanism}
\begin{lstlisting}[language=Python]
from visualize.auto_visualization import AutoVisualizer

# Get data for visualization
data_for_visualization = mywatermark.get_data_for_visualize(watermarked_image)

# Load Visualizer
visualizer = AutoVisualizer.load('GS', data_for_visualization=data_for_visualization)

# Draw diagrams on Matplotlib canvas
fig = visualizer.visualize(rows=2, cols=2, methods=['draw_watermark_bits', 'draw_reconstructed_watermark_bits', 'draw_inverted_latents', 'draw_inverted_latents_fft'])
\end{lstlisting}

\subsection{Cases for Evaluation}

\subsubsection{Watermark Detection Pipeline}
\begin{lstlisting}[language=Python]
# Dataset
my_dataset = StableDiffusionPromptsDataset(max_samples=200)
    
pipeline1 = WatermarkMediaDetectionPipeline(dataset=my_dataset, media_editor_list=[JPEGCompression(quality=60)],
show_progress=True, return_type=DetectionPipelineReturnType.SCORES) 

pipeline2 = UnWatermarkMediaDetectionPipeline(datmy_aset=my_dataset, media_editor_list=[],
show_progress=True, return_type=DetectionPipelineReturnType.SCORES)

detection_kwargs = {
    "num_inference_steps": 50,
    "guidance_scale": 1.0,
}

calculator = DynamicThresholdSuccessRateCalculator(labels=labels, rule=rules, target_fpr=target_fpr)
print(calculator.calculate(pipeline1.evaluate(my__watermark, detection_kwargs=detection_kwargs), pipeline2.evaluate(my__watermark, detection_kwargs=detection_kwargs))))
\end{lstlisting}
    
\subsubsection{Image Quality Analysis Pipeline}
\begin{lstlisting}[language=Python]
if metric == 'NIQE':
    my_dataset = StableDiffusionPromptsDataset(max_samples=max_samples)
    pipeline = DirectImageQualityAnalysisPipeline(dataset=my_dataset, 
        watermarkMed_iae_editor_list=[],
        unwatermarked_image_editor_list=[],
        analyzers=[NIQECalculator()],
        show_progress=True, 
        return_type=QualityPipelineReturnType.MEAN_SCORES)

elif metric == 'CLIP':
    my_dataset = MSCOCODataset(max_samples=max_samples)
    pipeline = ReferencedImageQualityAnalysisPipeline(dataset=my_dataset, 
        watermarked_image_editor_list=[],
        unwatermarked_image_editor_list=[],
        analyzers=[CLIPScoreCalculator()],
        unwatermarked_image_source='generated',
        reference_image_source='natural',
        show_progress=True, 
        return_type=QualityPipelineReturnType.MEAN_SCORES)
    
elif metric == 'FID':
    my_dataset = MSCOCODataset(max_samples=max_samples)
    pipeline = GroupImageQualityAnalysisPipeline(dataset=my_dataset, 
        watermarked_image_editor_list=[],
        unwatermarked_image_editor_list=[],
        analyzers=[FIDCalculator()],
        unwatermarked_image_source='generated',
        reference_image_source='natural',
        show_progress=True, 
        return_type=QualityPipelineReturnType.MEAN_SCORES)
elif metric == 'IS':
    my_dataset = StableDiffusionPromptsDataset(max_samples=max_samples)
    pipeline = GroupImageQualityAnalysisPipeline(dataset=my_dataset, 
        watermarked_image_editor_list=[],
        unwatermarked_image_editor_list=[],
        analyzers=[InceptionScoreCalculator()],
        show_progress=True, 
        return_type=QualityPipelineReturnType.MEAN_SCORES)

elif metric == 'LPIPS':
    my_dataset = StableDiffusionPromptsDataset(max_samples=10)
    pipeline = RepeatImageQualityAnalysisPipeline(dataset=my_dataset, 
        prompt_per_image=20,
        watermarked_image_editor_list=[],
        unwatermarked_image_editor_list=[],
        analyzers=[LPIPSAnalyzer()],
        show_progress=True, 
        return_type=QualityPipelineReturnType.MEAN_SCORES)

elif metric == 'PSNR':
    my_dataset = StableDiffusionPromptsDataset(max_samples=max_samples)
    pipeline = ComparedImageQualityAnalysisPipeline(dataset=my_dataset, 
        watermarked_image_editor_list=[],
        unwatermarked_image_editor_list=[],
        analyzers=[PSNRAnalyzer()],
        show_progress=True, 
        return_type=QualityPipelineReturnType.MEAN_SCORES)
else:
    raise ValueError('Invalid metric')


my_watermark = AutoWatermark.load(f'{algorithm_name}', 
    algorithm_config=f'config/{algorithm_name}.json',
    diffusion_config=diffusion_config)
print(pipeline.evaluate(my_watermark))
\end{lstlisting}

\subsubsection{Video Quality Analysis Pipeline}
\begin{lstlisting}[language=Python]
from evaluation.tools.video_quality_analyzer import (
    SubjectConsistencyAnalyzer,
    MotionSmoothnessAnalyzer, 
    DynamicDegreeAnalyzer,
    BackgroundConsistencyAnalyzer,
    ImagingQualityAnalyzer
)

# Load VBench dataset
my_dataset = VBenchDataset(max_samples=200, dimension=dimension)

# Initialize analyzer based on metric
if metric == 'subject_consistency':
    analyzer = SubjectConsistencyAnalyzer(device=device)
elif metric == 'motion_smoothness':
    analyzer = MotionSmoothnessAnalyzer(device=device)
elif metric == 'dynamic_degree':
    analyzer = DynamicDegreeAnalyzer(device=device)
elif metric == 'background_consistency':
    analyzer = BackgroundConsistencyAnalyzer(device=device)
elif metric == 'imaging_quality':
    analyzer = ImagingQualityAnalyzer(device=device)
else:
    raise ValueError(f'Invalid metric: {metric}. Supported metrics: subject_consistency, motion_smoothness, dynamic_degree, background_consistency, imaging_quality')

# Create video quality analysis pipeline
pipeline = DirectVideoQualityAnalysisPipeline(
    dataset=my_dataset,
    watermarked_video_editor_list=[],
    unwatermarked_video_editor_list=[],
    watermarked_frame_editor_list=[],
    unwatermarked_frame_editor_list=[],
    analyzers=[analyzer],
    show_progress=True,
    return_type=QualityPipelineReturnType.MEAN_SCORES
)

print(pipeline.evaluate(my_watermark))
\end{lstlisting}
  \section{Details of the Integrated Watermarking Algorithms}
\label{appendix:algorithms}
Table~\ref{tab:watermarking_algorithms} presents eight watermarking algorithms
integrated into \textsc{MarkDiffusion}. These algorithms comprehensively
represent the two major paradigms of generative watermarking, covering both image
and video generation tasks. Each algorithm features distinct design objectives
and unique innovations.

\begin{table}[t]

  \caption{Details of watermarking algorithms implemented in our toolkit.}
  \centering
  \begin{tabular}{p{5cm} p{2cm} p{7cm}}
   \toprule
    \textbf{Algorithm Name}            & \textbf{Category} & \textbf{Methodology}                                                                                                                                                                                                       \\
    \midrule
    \small Tree-Ring \citep{wenTreeRingWatermarksFingerprints2023}        & \small Pattern-Based & \small Embeds a Fourier concentric ring pattern into the initial latent space, with detection achieved by inverting the generated image back to its latent representation and calculating the distance to the original pattern. \\
    \small Ring-ID \citep{ciRingIDRethinkingTreeRing2024}          & \small Pattern-Based & \small Base on Tree-Ring, imprints multiple keys into different channels, significantly enhances watermark verification and multi-key identification.                                                                           \\
    \small ROBIN \citep{huangROBINRobustInvisible2025}            & \small Pattern-Based & \small Base on Tree-Ring, embeds a watermark pattern into an intermediate diffusion state and produce an optimized guiding signal during generation, in order to minimize the impact on image semantics.                                                                                   \\
    \small WIND \citep{gunnUndetectableWatermarkGenerative2024}             & \small Pattern-Based & \small Enable a large set of keys by maintaining a large set of initial noises with different seed and dividing them into different group with Fourier patterns imprinted into them as group identifiers.                                                                                                                                                 \\
    \small Gaussian-Shading \citep{yangGaussianShadingProvable2024} & \small Key-Based & \small Chooses a fixed quadrant of latent space as the watermarking key, repeats the key to create a distortion-free initial latent and generate images from that latent.                                                       \\
    \small PRC-Watermark \citep{christPseudorandomErrorCorrectingCodes2024}    & \small Key-Based & \small Leverages pseudo-random error-correcting codes to create the initial latents for generation, achieves computational undetectability.                                                                                     \\
    \small SEAL \citep{arabiSEALSemanticAware2025}             & \small Key-Based & \small Correlates the watermark directly with the image's semantics, uses the semantic vector of the unwatermarked images and the secret salt to create the initial noise, independent of access to historical databases of watermarking keys.                                                                                  \\
    \small VideoShield \citep{huVideoShieldRegulatingDiffusionbased2025}      & \small Key-Based & \small Extend the core idea of Gaussian Shading to the video generation domain.                                                                                                                                                 \\
    \bottomrule
  \end{tabular}
  \label{tab:watermarking_algorithms}
  \vspace{-0.5cm}
\end{table}
  \section{Additional Visualization Examples}
\label{appendix:visualize}
This section provides visualization examples for each supported watermarking algorithm, along with invocation code.

\noindent
\textbf{Tree-Ring Visualization:} 
\begin{lstlisting}[language=Python]
from visualize.auto_visualization import AutoVisualizer

data_for_visualization = mywatermark.get_data_for_visualize(watermarked_image)

visualizer = AutoVisualizer.load('TR', data_for_visualization=data_for_visualization)

method_kwargs = [{},{"channel": 0}, {}, {"channel": 0}, {}]

fig = visualizer.visualize(
    rows=1, 
    cols=5, 
    methods=['draw_pattern_fft','draw_orig_latents_fft', 'draw_watermarked_image', 'draw_inverted_latents_fft', 'draw_inverted_pattern_fft'], 
    method_kwargs=method_kwargs, 
    save_path='TR_watermark_visualization.pdf'
    )
\end{lstlisting}
\begin{figure}[H]
    \centering
    \includegraphics[width=1\textwidth]{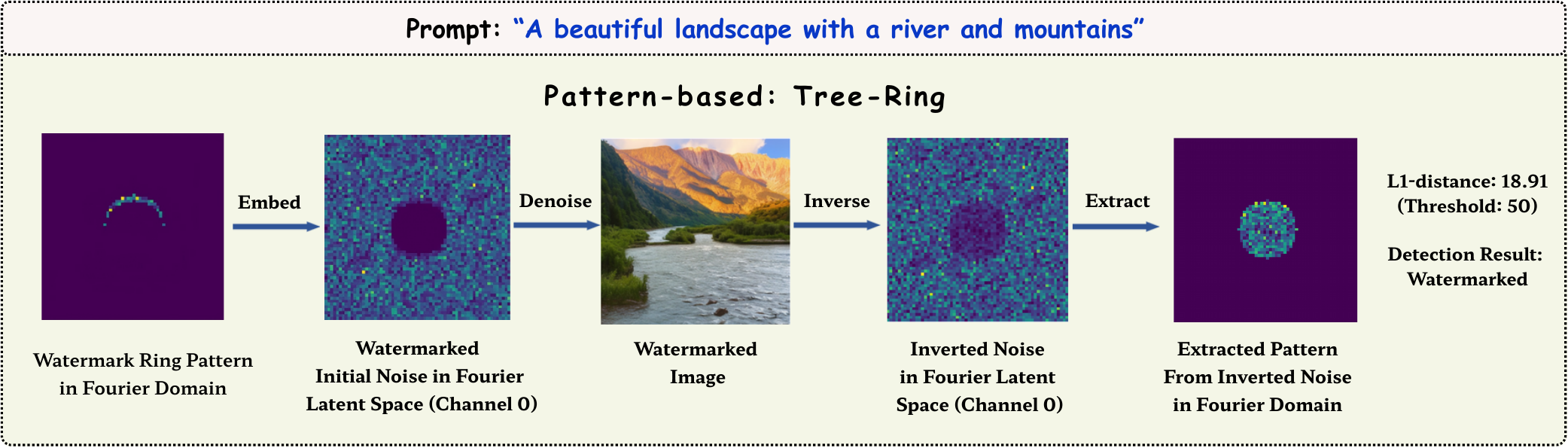}
    \caption{A visualization example of Tree-Ring watermark.}
    \label{fig:tree_ring_vis}
\end{figure}

\noindent
\textbf{Gaussian-Shading Visualization:}
\begin{lstlisting}[language=Python]
from visualize.auto_visualization import AutoVisualizer

data_for_visualization = mywatermark.get_data_for_visualize(watermarked_image)

visualizer = AutoVisualizer.load('GS', data_for_visualization=data_for_visualization)

method_kwargs = [{"channel": 0},{"channel": 0}, {}, {"channel": 0}, {"channel": 0}]

fig = visualizer.visualize(
    rows=1, 
    cols=5, 
    methods=[ 'draw_watermark_bits', 'draw_orig_latents', 'draw_watermarked_image','draw_inverted_latents', 'draw_reconstructed_watermark_bits'], 
    method_kwargs=method_kwargs, 
    save_path='GS_watermark_visualization.pdf'
    )
\end{lstlisting}
\begin{figure}[H]
    \centering
    \includegraphics[width=1\textwidth]{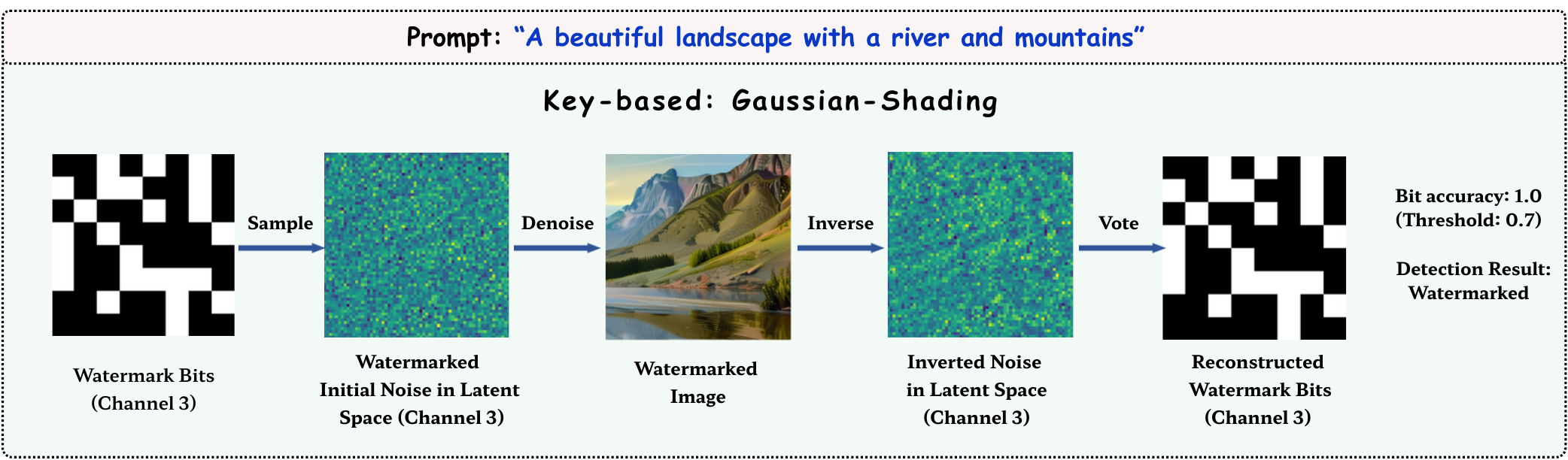}
    \caption{A visualization example of Gaussian-Shading watermark.}
    \label{fig:gause_shading_vis}
\end{figure}
\noindent
\textbf{RingID Visualization:}
\begin{lstlisting}[language=Python]
from visualize.auto_visualization import AutoVisualizer

data_for_visualization = mywatermark.get_data_for_visualize(watermarked_image)

visualizer = AutoVisualizer.load('RI', data_for_visualization=data_for_visualization)

method_kwargs = [{}, {"channel": 3}, {}, {"channel": 3}, {}]
fig = visualizer.visualize(
    rows=1, 
    cols=5, 
    methods=['draw_ring_pattern_fft',
    'draw_orig_latents_fft', 'draw_watermarked_image',
    'draw_inverted_latents_fft',
    'draw_heter_pattern_fft'], 
    method_kwargs=method_kwargs, 
    save_path='RI_watermark_visualization.pdf'
)
\end{lstlisting}
\begin{figure}[H]
    \centering
    \includegraphics[width=1\textwidth]{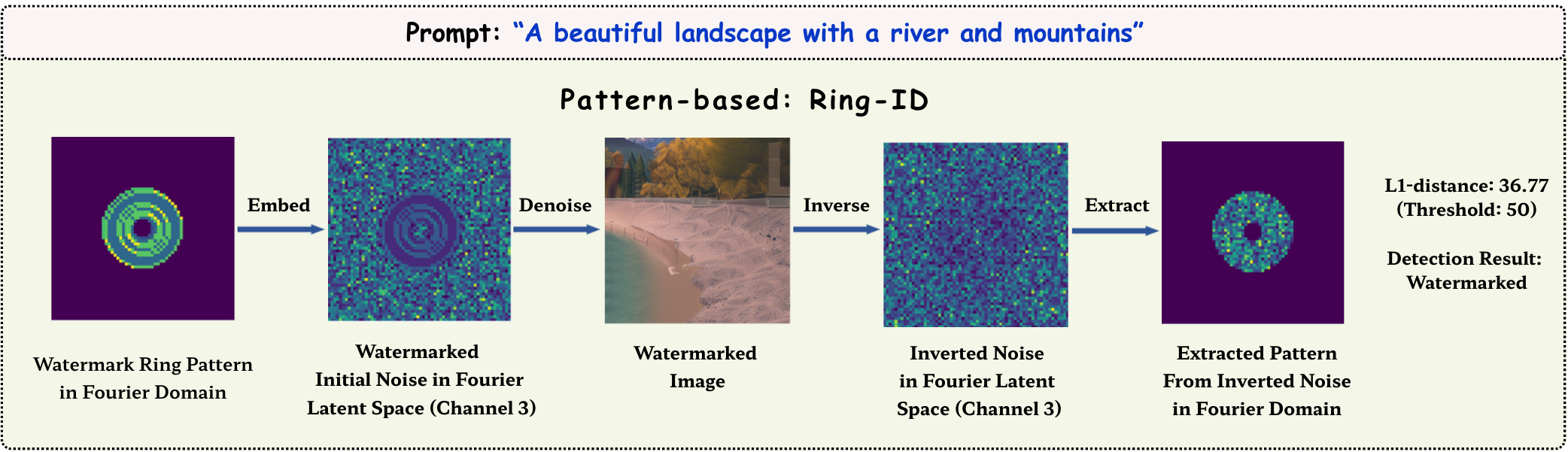}
    \caption{A visualization example of RingID watermark.}
    \label{fig:ringid_vis}
\end{figure}
\noindent
\textbf{SEAL Visualization:}
\begin{lstlisting}[language=Python]
from visualize.auto_visualization import AutoVisualizer
data_for_visualization = mywatermark.get_data_for_visualize(watermarked_image)

visualizer = AutoVisualizer.load('SEAL', data_for_visualization=data_for_visualization)

method_kwargs = [{}, {"channel": 2}, {}, {"channel": 2}, {}]

fig = visualizer.visualize(
    rows=1, 
    cols=5, 
    methods=['draw_embedding_distributions','draw_orig_latents', 'draw_watermarked_image', 'draw_inverted_latents', 'draw_patch_diff'], 
    method_kwargs=method_kwargs, 
    save_path='SEAL_watermark_visualization.pdf'
    )
\end{lstlisting}
\begin{figure}[H]
    \centering
    \includegraphics[width=1\textwidth]{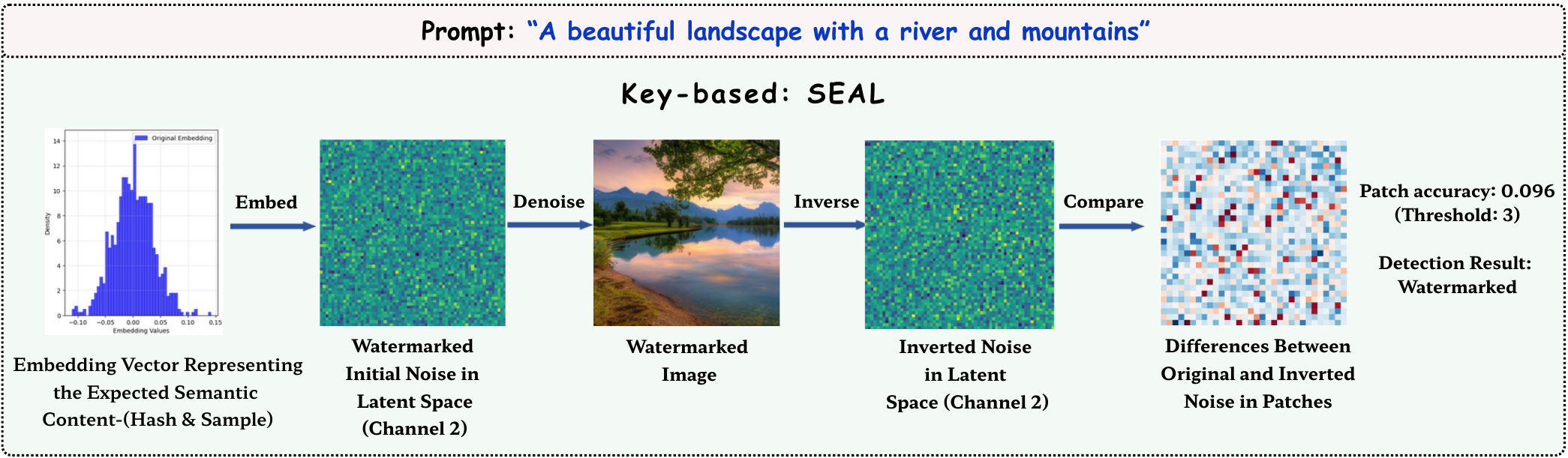}
    \caption{A visualization example of SEAL watermark.}
    \label{fig:seal_vis}
\end{figure}
\noindent
\textbf{WIND Visualization:} 
\begin{lstlisting}[language=Python]
from visualize.auto_visualization import AutoVisualizer

data_for_visualization = mywatermark.get_data_for_visualize(watermarked_image)

visualizer = AutoVisualizer.load('WIND', data_for_visualization=data_for_visualization)

method_kwargs = [{"channel": 2}, {"channel": 2}, {}, {"channel": 2}, {"channel": 2}]

fig = visualizer.visualize(
    rows=1, 
    cols=5, 
    methods=['draw_group_pattern_fft', 'draw_orig_latents_fft', 'draw_watermarked_image', 'draw_inverted_latents_fft', 'draw_inverted_group_pattern_fft'], 
    method_kwargs=method_kwargs, 
    save_path='WIND_watermark_visualization.pdf'
    )
\end{lstlisting}
\begin{figure}[H]
    \centering
    \includegraphics[width=1\textwidth]{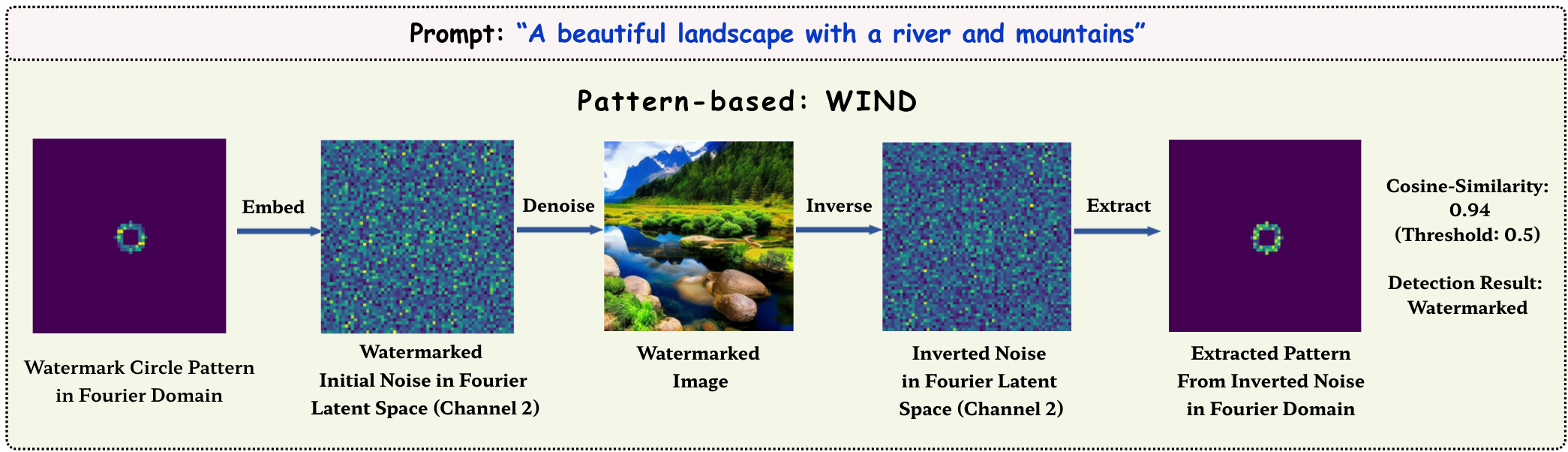}
    \caption{A visualization example of WIND watermark.}
    \label{fig:wind_vis}
\end{figure}
\noindent
\textbf{ROBIN Visualization:}
\begin{lstlisting}[language=Python]
from visualize.auto_visualization import AutoVisualizer

data_for_visualization = mywatermark.get_data_for_visualize(watermarked_image)

visualizer = AutoVisualizer.load('ROBIN', data_for_visualization=data_for_visualization)

method_kwargs = [{}, {"channel": 3}, {}, {"channel": 3}, {}]

fig = visualizer.visualize(
    rows=1, 
    cols=5, 
    methods=['draw_pattern_fft', 'draw_orig_latents_fft', 'draw_watermarked_image', 'draw_inverted_latents_fft', 'draw_inverted_pattern_fft'], 
    method_kwargs=method_kwargs, 
    save_path='ROBIN_watermark_visualization.pdf'
    )
\end{lstlisting}

\begin{figure}[H]
    \centering
    \includegraphics[width=1\textwidth]{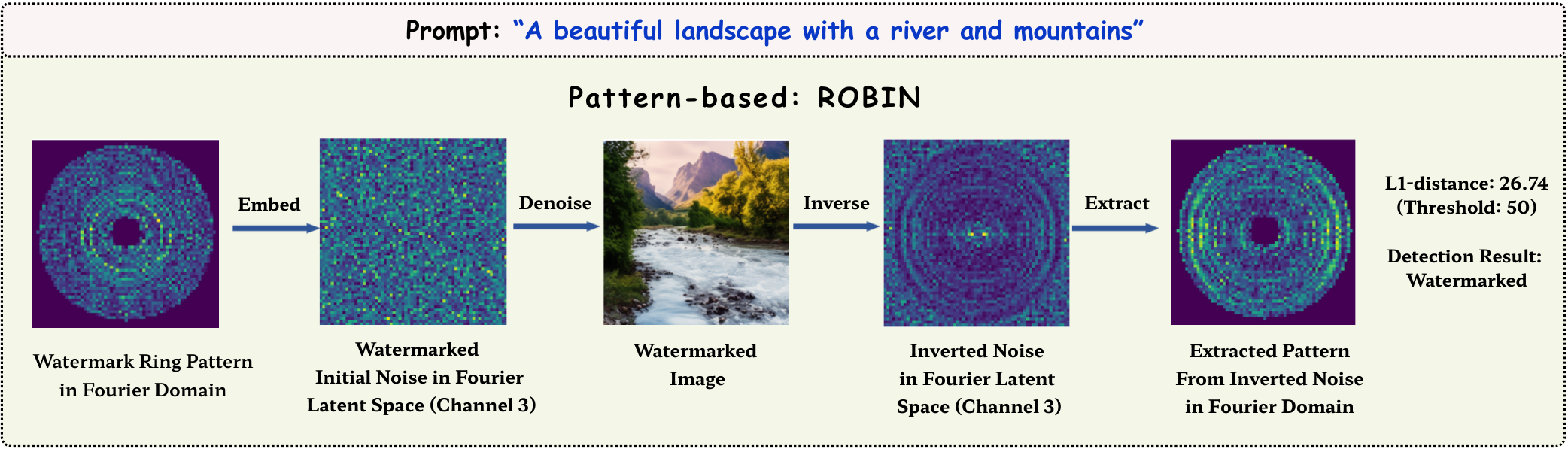}
    \caption{A visualization example of ROBIN watermark.}
    \label{fig:robin_vis}
\end{figure}
\noindent
\textbf{PRC Visualization:}
\begin{lstlisting}[language=Python]
from visualize.auto_visualization import AutoVisualizer

data_for_visualization = mywatermark.get_data_for_visualize(watermarked_image)

visualizer = AutoVisualizer.load('PRC', data_for_visualization=data_for_visualization)

method_kwargs = [{}, {"channel": 3}, {}, {"channel": 3}, {}]

fig = visualizer.visualize(
    rows=1, 
    cols=5, 
    methods=['draw_codeword','draw_orig_latents', 'draw_watermarked_image', 'draw_inverted_latents','draw_recovered_codeword'], 
    method_kwargs=method_kwargs, 
    save_path='PRC_watermark_visualization.pdf'
    )
\end{lstlisting}
\begin{figure}[H]
    \centering
    \includegraphics[width=1\textwidth]{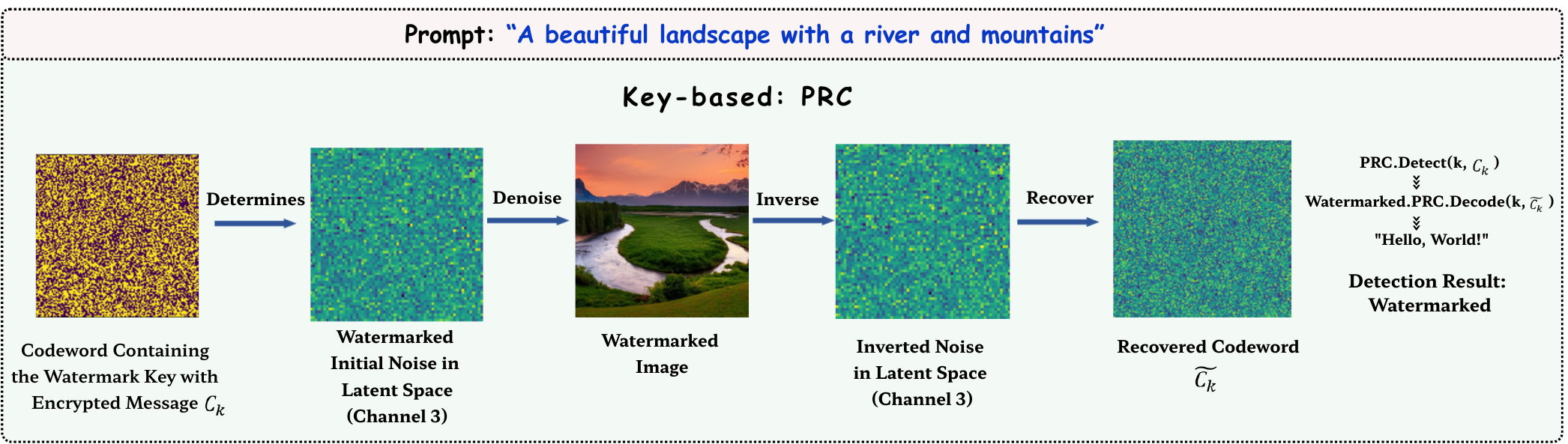}
    \caption{A visualization example of PRC watermark.}
    \label{fig:prc_vis}
\end{figure}
\textbf{VideoShield Visualization:}
\begin{lstlisting}[language=Python]
from visualize.auto_visualization import AutoVisualizer

data_for_visualization = mywatermark.get_data_for_visualize(
    video_frames=watermarked_video,
    prompt="",
    guidance_scale=1.0,
    num_inference_steps=25,
)
visualizer = AutoVisualizer.load('VideoShield', data_for_visualization=data_for_visualization)

method_kwargs = [{"channel": 1}, {"channel": 1}, {"num_frames": 4}, {"channel": 1}, {"channel": 1}]

fig = visualizer.visualize(
    rows=1, 
    cols=5, 
    methods=['draw_watermark_bits','draw_orig_latents', 'draw_watermarked_video_frames', 'draw_inverted_latents', 'draw_reconstructed_watermark_bits'], 
    method_kwargs=method_kwargs, 
    save_path='VideoShield_watermark_visualization.pdf'
    )
\end{lstlisting}
\begin{figure}[H]
    \centering
    \includegraphics[width=1.0\textwidth]{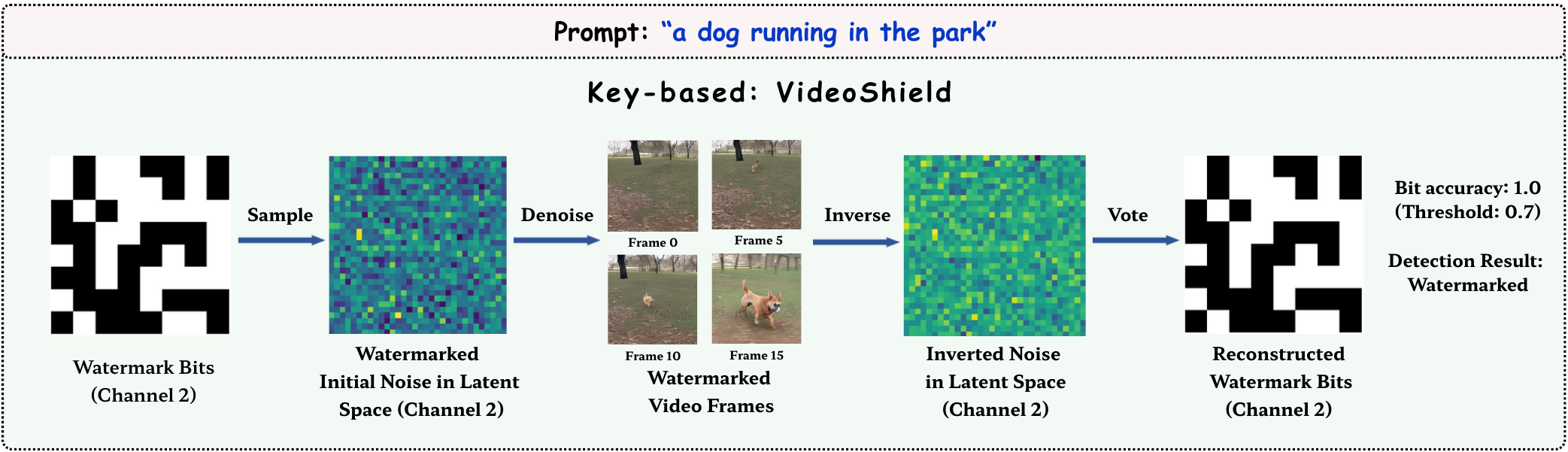}
    \caption{A visualization example of VideoShield watermark.}
    \label{fig:prc_vis}
\end{figure}
\end{document}